\documentclass[prl,aps,amssymb,showpacs]{revtex4}
\usepackage{graphicx}
\begin{document}

\title{Asymmetric Fermion Superfluid with Inter- and Intra-Species Pairings}
\author{Xuguang Huang, Xuewen Hao and Pengfei Zhuang }
\affiliation{Physics Department, Tsinghua University, Beijing
100084, China}
\date{\today}
\begin{abstract}
We investigate the phase structure of an asymmetric fermion
superfluid with inter- and intra-species pairings. The
introduction of the intra-species pairing mechanism in canonical
ensemble changes significantly the phase diagram and brings in a
new state with coexisting inter- and intra-species pairings.
Different from the case with only inter-species pairing, all the
fermion excitations are fully gapped in the region with
intra-species pairing.
\end{abstract}

\maketitle

Recently, the fermion pairing between different species with
mismatched Fermi surfaces prompted great interest in both
experimental\cite{zwie,part,shin} and
theoretical\cite{son,calson,he1,iskin,bulgac,ho,giannakis,hu,machida1,caldas}
studies. In conventional fermion superfluid the ground state is well
described by the BCS theory, while for asymmetric fermion superfluid
the phase structure is much more rich and the pairing mechanism is
not yet very clear. Various exotic phases have been suggested, such
as the Sarma phase\cite{sarma} or gapless phase\cite{huang1} where
the superfluid component is breached by a normal component in
momentum space\cite{liu}, the Fulde-Ferrel-Larkin-Ovchinnikov(FFLO)
phase\cite{fuld} where the Cooper pair is momentum dependent, the
phase with deformed Fermi surfaces where the two Fermi surfaces
cross to each other and the total momentum zero Cooper pairs could
form at the crossing node\cite{muth,sedr}, and the phase
separation\cite{beda,cald} in coordinate space where the normal and
superfluid components are inhomogeneously mixed.

For many fermion superfluid systems, the fermions from the same
species can form Cooper pairs as well. In ultracold atom gas like
$^6$Li and $^{40}$K, there exist pairings between different
elements and between different states of the same element. In
color superconducting phase of dense quark matter\cite{huang}, the
quarks of different flavors can form total spin zero Cooper pairs,
and the quarks of the same flavor can be combined into total spin
one pairs which describe better the cooling rates of quark
stars\cite{schmit}. In neutron stars\cite{lomb}, proton-proton,
neutron-neutron and neutron-proton pairings are all possible.
Recently discovered high temperature superconductivity of
MgB$_2$\cite{iava,bouquet,tsuda,geek} can also be well described
by an extended two-band BCS theory where the electrons from the
same energy band form Cooper pairs.

Since the ratio of the intra- to inter-species pairing gap is
already $10\%$ for color superconductivity and can even reach
$50\%$ for nuclear superfluidity and ultracold atom gas, the
competition between the two types of pairings should be strong.
Considering that the pairing between fermions from the same
species happens at the common Fermi surface, the intra-species
pairing will become more favored than the inter-species pairing,
when the Fermi surface mismatch between the two kinds of species
is large enough. Therefore, some familiar phenomena in the
superfluid with only inter-species pairing may be washed out by
the introduction of intra-species pairing, and some new phases may
arise. For instance, the interesting gapless state may disappear,
the inhomogeneous FFLO state at low
temperature\cite{machida1,he2,yan} may be eaten up by the
homogeneous intra-species pairing state, and the mixed state of
BCS superfluid and normal phase\cite{beda,cald} may be replaced by
a new phase where the inter- and intra- species pairings coexist.
In this paper we propose a general model to investigate the
competition between the inter- and intra-species pairings.

We consider a fermion fluid containing two species $a$ and $b$ with
masses $m_a$ and $m_b$ and chemical potentials $\mu_a$ and $\mu_b$.
The system can be described by the Lagrangian density
\begin{equation}
\label{h1} {\cal
L}=\sum_{i,\sigma}\bar{\psi}_i^\sigma\left[-\frac{\partial}{\partial\tau}+\frac{\nabla^2}{2m_i}+\mu_i\right]\psi_i^\sigma
-\frac{g}{2}\sum_{\sigma\ne\sigma',\rho\ne\rho'}\bar{\psi}_a^\sigma\bar{\psi}_b^{\sigma'}\psi_b^{{\rho'}}\psi_a^{\rho}
-\sum_i
g_i\bar{\psi}_i^\uparrow\bar{\psi}_i^\downarrow\psi_i^\downarrow\psi_i^\uparrow,
\end{equation}
where $\psi_i^\sigma(x)$ are fermion fields with $i=a,b$ and
(pseudo-)spin $\sigma=\uparrow,\downarrow$, the coupling constants
$g, g_a$ and $g_b$ controlling respectively the inter- and
intra-species pairings, are negative to keep the interactions
attractive. The Lagrangian has the symmetry $U_a(1)\otimes U_b(1)$
with the element $U(\theta_a,\theta_b)$ defined as
$U(\theta_a,\theta_b)\psi_a^\sigma=e^{i\theta_a}\psi_a^\sigma$ and $
U(\theta_a,\theta_b)\psi_b^\sigma=e^{i\theta_b}\psi_b^\sigma$.

We introduce the condensates of $a$-$a$, $b$-$b$ and $a$-$b$ pairs,
$\Phi_a=-g_a\langle\psi_a^\downarrow\psi_a^\uparrow\rangle,
\Phi_b=-g_b\langle\psi_b^\downarrow\psi_b^\uparrow\rangle$ and
$\Phi=-(g/2)\sum_{\sigma\ne\sigma'}\langle\psi_b^{\sigma'}\psi_a^\sigma\rangle$.
To incorporate the FFLO state in the study, we assume that the
condensates are in the form of $\Phi_a({\bf x})=\Delta_a e^{2i{\bf
q}_a\cdot{\bf x}}, \Phi_b({\bf x})=\Delta_b e^{2i{\bf q}_b\cdot{\bf
x}}$ and $\Phi({\bf x})=\Delta e^{i({\bf q}_a+{\bf q}_b)\cdot{\bf
x}}$ with $\Delta_a, \Delta_b$ and $\Delta$ being independent of
${\bf x}$. Obviously, the translational symmetry and rotational
symmetry in the FFLO state are spontaneously broken. Note that, for
the sake of simplicity, the FFLO state we considered here is its
simplest pattern, namely the single plane wave FFLO state or the
so-called FF state.

The key quantity of a thermodynamic system is the partition
function ${\cal Z}$ which can be represented by the path integral
\begin{equation}
{\cal Z}=\prod\int{\cal D}\psi_i^\sigma{\cal
D}\bar{\psi}_i^\sigma\exp{(\int d\tau d{\bf x}{\cal L})}\ .
\end{equation}
By performing a gauge transformation for the fermion fields,
$\chi_a^\sigma=e^{-i{\bf q}_a\cdot{\bf x}}\psi_a^\sigma$ and
$\chi_b^\sigma=e^{-i{\bf q}_b\cdot{\bf x}}\psi_b^\sigma$, the path
integral over $\chi_{a,b}^\sigma$ in the partition function ${\cal
Z}$ of the system at mean field level can be calculated easily,
and we obtain the mean field thermodynamic potential
\begin{eqnarray}
\label{omega}
\Omega &=& -{T\over V}\ln{\cal Z}\nonumber\\
&=&-{\Delta_a^2\over g_a}-{\Delta_b^2\over g_b}-{2\Delta^2\over
g}-T\sum_{n,{\bf p}}{\rm Tr}\ln G^{-1}( i\omega_n, {\bf p})
\end{eqnarray}
in the imaginary time formulism of finite temperature field
theory, where $\omega_n=(2n+1)\pi T$ is the fermion frequency,
${\bf p}$ is the fermion momentum, and the inverse of the
Nambu-Gorkov propagator can be written as
\begin{equation}
G^{-1}=\left(\begin{array}{cccc}i\omega_n-\epsilon_a^+ & 0 & \Delta & \Delta_a\\
0 & i\omega_n-\epsilon_b^+ & \Delta_b & -\Delta\\
\Delta^* & \Delta_b^* & i\omega_n+\epsilon_b^- & 0\\
\Delta_a^* & -\Delta^* & 0 &i\omega_n+\epsilon_a^-
\end{array}\right)
\end{equation}
with $\epsilon_i^\pm=({\bf p}\pm{\bf q}_i)^2/(2m_i)-\mu_i$.

The Lagrangian (\ref{h1}) is non-renormalizable, a regularization
scheme should be applied. We use the s-wave scattering lengthe
regularization which relates the bare coupling constants $g,g_a$
and $g_b$ to the low energy limit of the corresponding two-body
T-matrices in vacuum by
\begin{eqnarray}
\frac{m}{4\pi a_a}&=&\frac{1}{g_a}+\sum_{\bf p}\frac{m}{{\bf p}^2},\nonumber\\
\frac{m}{4\pi a_b}&=&\frac{1}{g_b}+\sum_{\bf p}\frac{m}{{\bf p}^2},\nonumber\\
\frac{m}{4\pi a}&=&\frac{1}{g}+\sum_{\bf p}\frac{m}{{\bf p}^2}
\end{eqnarray}
with the s-wave scattering lengthes $a_a,a_b$ and $a$. Such a
scheme is reliable in the whole region of interacting strength,
and therefore it is possible to extend our study to the BCS-BEC
crossover, although we focus in this paper only on the weak
coupling BCS region. For simplicity, we have taken the same
particle masses $m_a=m_b=m$ and will take all the condensates as
real numbers.

One should note that if the relative momentum ${\bf q}_a-{\bf
q}_b$ is large enough, the uniform superfluid would be unstable
due to the stratification of the superfluid components
characterized by $\Phi_a({\bf x})$ and $\Phi_b({\bf x})$. The case
here is analogous to the multi-component BEC. To avoid such a
dynamic instability\cite{khal,mine,nepo}, we choose ${\bf
q}_a={\bf q}_b={\bf q}$\cite{note}. After computing the frequency
summation and taking a Bogoliubov-Valatin transformation from
particles $a$ and $b$ to quasi-particles, the thermodynamic
potential can be expressed in terms of the quasi-particles,
\begin{equation}
\label{omega} \Omega = -{m\Delta_a^2\over {4\pi
a_a}}-{m\Delta_b^2\over {4\pi a_b}}-{2m\Delta^2\over {4\pi
a}}+(\Delta_a^2+\Delta_b^2+2\Delta^2)\sum_{\bf p}{m\over {\bf
p}^2}+\sum_{\bf p}\sum_{i=a,b} \epsilon_i^- -\sum_{\bf
p}\sum_{j,k=\pm}\left[{E_j^k\over
2}+T\ln\left(1+e^{-E_j^k/T}\right)\right],
\end{equation}
where $E_\pm^\mp$ are the quasi-particle energies
\begin{equation}
E_\pm^\mp=\sqrt{\epsilon_+^2+\delta\epsilon^2\pm\sqrt{\epsilon_-^4+\epsilon_\Delta^4}}\mp\delta\epsilon
\end{equation}
with $\epsilon_\pm, \delta\epsilon$ and $\epsilon_\Delta$ defined
as
\begin{eqnarray}
&& \epsilon_\pm^2
= \left[\left(\epsilon_a^+\epsilon_a^-+\Delta_a^2+\Delta^2\right)\pm\left(\epsilon_b^+\epsilon_b^-+\Delta_b^2+\Delta^2\right)\right]/2,\nonumber\\
&& \delta\epsilon =
(\epsilon_a^+-\epsilon_a^-)/2=(\epsilon_b^+-\epsilon_b^-)/2,\nonumber\\
&& \epsilon_\Delta^4 =
\Delta^2[(\epsilon_a^+-\epsilon_b^+)(\epsilon_a^--\epsilon_b^-)+(\Delta_b-\Delta_a)^2].
\end{eqnarray}
$\Delta_a\neq 0$ and $\Delta_b\neq 0$ correspond, respectively, to
the spontaneous symmetry breaking patterns $U_a(1)\otimes
U_b(1)\rightarrow U_b(1)$ and $U_a(1)$, and $\Delta\neq 0$ means the
breaking pattern $U_a(1)\otimes U_b(1)\rightarrow U_{a-b}(1)$ with
the element $U(\theta)$ defined as
$U(\theta)\psi_a^\sigma=e^{i\theta}\psi_a^\sigma$ and
$U(\theta)\psi_b^\sigma=e^{-i\theta}\psi_b^\sigma$.

The condensates and the FFLO momentum as functions of temperature
and chemical potentials are determined by the gap equations,
\begin{equation}
\label{gap1}
{\partial\Omega\over \partial\Delta_a}=0,\ \
{\partial\Omega\over \partial\Delta_b}=0,\ \
{\partial\Omega\over
\partial\Delta}=0,\ \
{\partial\Omega\over \partial{\bf q}}=0,
\end{equation}
and the ground state of the system is specified by the minimum of
the thermodynamic potential, namely by the second-order
derivatives.

To see clearly the effect of the mismatch between the two species
$a$ and $b$, we introduce the average chemical potential
$\mu=(\mu_a+\mu_b)/2$ and the chemical potential mismatch
$\delta\mu=(\mu_b-\mu_a)/2$ instead of $\mu_a$ and $\mu_b$.
Without loss of generality, we assume $\mu_b>\mu_a$. We choose
$p_F a, p_F a_a, p_F a_b$ as the free parameters of the model,
where $p_F=\sqrt{2m\mu}$ is the average Fermi momentum, and set
$p_F a=-0.58$ and $p_F a_a=p_F a_b=0.73\ p_F a$ in the numerical
calculation. We have checked that in the BCS region of $0<p_F
|a|,\ p_F |a_a|,\ p_F |a_b| <1$ and $0<|a_a|=|a_b|<|a|$ to
guarantee weak interaction for $a$-$b$ pairing and more weak
coupling for $a$-$a$ and $b$-$b$ pairings, there is no qualitative
change in the obtained phase diagrams. Note that, in the symmetric
case with $\delta\mu=0$, from a direct integration of the gap
equation, one obtains the famous result\cite{delta0} $\Delta_0
\simeq 8\mu e^{-2}\exp{\left[-\pi/(2p_F |a|)\right]}$ for the
inter-species pairing gap $\Delta_0$ at zero temperature.

\begin{figure}[!htb]
\begin{center}
\includegraphics[width=8cm]{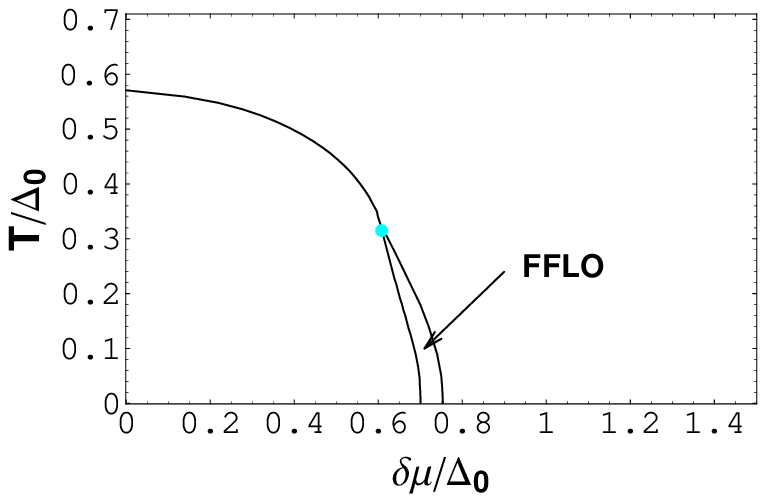}
\includegraphics[width=8cm]{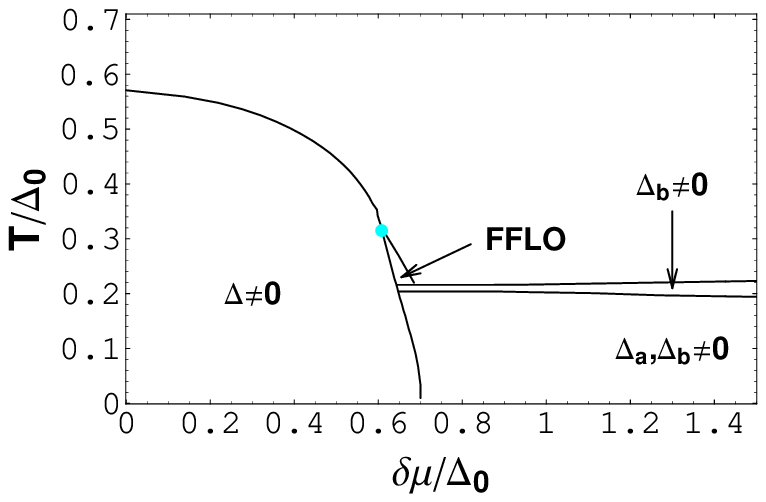}
\caption{ The phase diagram of an asymmetric fermion superfluid in
grand canonical ensemble. $\delta\mu$ is the chemical potential
mismatch between the two species, and the average chemical potential
is fixed to be $\mu=50\Delta_0$ with $\Delta_0$ being the
corresponding symmetric gap at $\delta\mu=0$. The left panel is for
the familiar case with only inter-species pairing, and in the right
panel the intra-species pairing is included as well. } \label{fig1}
\end{center}
\end{figure}

We first consider systems of grand canonical ensemble with fixed
chemical potentials. The phase diagram in $T-\delta\mu$ plane is
shown in Fig.\ref{fig1}. The left panel is the familiar case without
intra-species pairing\cite{taka}. Both thermal fluctuations and
large chemical potential mismatch can break the Cooper pairs. The
homogeneous BCS state can exist at low temperature and low mismatch,
and the inhomogeneous FFLO state survives only in a narrow mismatch
window. The phase transition from the superfluid to normal phase is
of second order, and the transition from the homogeneous to
inhomogeneous superfluid at zero temperature is of first order and
happens at $\delta\mu_c=\Delta_0/\sqrt{2}$ \cite{taka}. When the
intra-species pairing is included as well, see the right panel of
Fig.\ref{fig1}, the inhomogeneous FFLO state of $a$-$b$ pairing is
eaten up by the homogeneous superfluid of $a$-$a$ and $b$-$b$
pairings at low temperature, just as we expected, and survives only
in a small triangle at high temperature. The phase transition from
the $a$-$b$ pairing superfluid to the $a$-$a$ and $b$-$b$ pairing
superfluid is of first order. From the assumption of $\mu_b
> \mu_a$, the temperature to melt the condensate $\Delta_b$ is higher than that to melt $\Delta_a$,
which leads to a phase with only $b$-$b$ pairing. Since $\mu_b$
increases and $\mu_a$ decreases with mismatch $\delta\mu$, the
region with only $b$-$b$ pairing becomes more and more wide when
$\delta\mu$ increases. Note that, for systems with fixed chemical
potentials there is no mixed phase of inter- and intra-species
pairings, and the situation is similar to a three-component
fermion system\cite{paan}.

For many physical systems, the fixed quantities are not chemical
potentials $\mu_a$ and $\mu_b$ but particle number densities
$n_a=-\partial\Omega/\partial\mu_a$ and
$n_b=-\partial\Omega/\partial\mu_b$, or equivalently speaking the
total number density $n=n_b+n_a$ and the number density asymmetry
$\alpha=\delta n/n$ with $\delta n = n_b-n_a$. Such systems are
normally described by the canonical ensemble and the essential
quantity is the free energy which is related to the thermodynamic
potential by a Legendre transformation,
$\mathcal{F}=\Omega+\mu_an_a+\mu_bn_b=\Omega+\mu n+\delta\mu\delta
n$.

It is easy to prove that the gap equations in the canonical
ensemble are equivalent to (\ref{gap1}) in the grand canonical
ensemble and give the same solutions. While the candidates of the
ground state of the system are the same in both ensembles, the
stability conditions for the two ensembles are very different and
lead to much more rich phase structure in systems with fixed
number densities. Our method to obtain the phase diagram at fixed
total number density $n$ is as the following. We first calculate
the gap equations (\ref{gap1}) and obtain all the possible
homogeneous phases and inhomogeneous FFLO phase, then compare
their free energies to extract the lowest one at fixed $T$ and
$\alpha$, and finally investigate the stability of the system
against number fluctuations by computing the number susceptibility
matrix
\begin{equation}
\chi={\partial^2\mathcal{F}\over \partial n_i\partial
n_j}=-{\partial^2\Omega\over\partial\mu_i\partial\mu_j} +Y_i
R^{-1}Y_j^\dag,
\end{equation}
where $Y_i$ are susceptibility vectors with elements
$\left(Y_i\right)_m =
\partial^2\Omega/(\partial\mu_i\partial x_m)$ and $R$
is a susceptibility matrix with elements
$R_{mn}=\partial^2\Omega/(\partial x_m\partial x_n)$ in the order
parameter space constructed by $x=(\Delta_a, \Delta_b, \Delta, {\bf
q})$. The state with non positive-definite $\chi$ (denoted by
$\chi<0$) is unstable against number fluctuations and may be a phase
separation\cite{chi}.

\begin{figure}[!htb]
\begin{center}
\includegraphics[width=8cm]{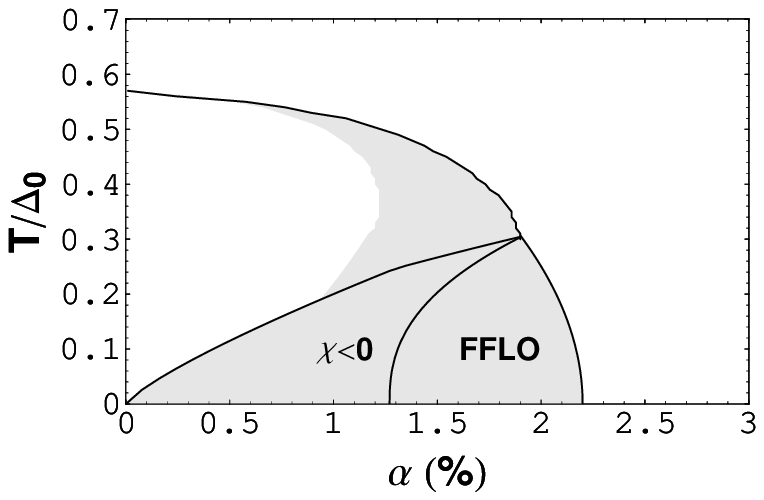}
\includegraphics[width=8cm]{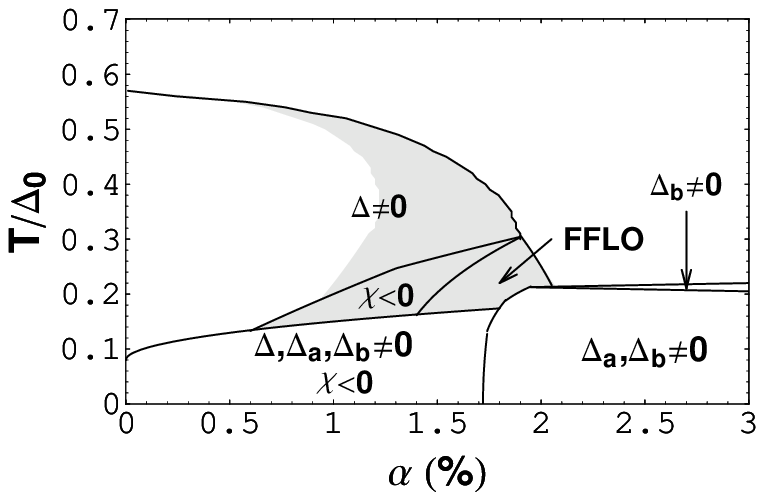}
\caption{ The phase diagram of a mismatched fermion superfluid in
canonical ensemble. $\alpha$ describes the number asymmetry between
the two species, and the total number $n$ is fixed corresponding to
$\mu=50\Delta_0$. The left panel is for the familiar case with only
inter-species pairing. In the right panel the intra-species pairing
is included as well. The shadowed regions indicate the gapless
superfluid.} \label{fig2}
\end{center}
\end{figure}

The phase diagram without intra-species pairing can be found in
\cite{machida1,he2,yan}. As a comparison we recalculate it and show
it in the left panel of Fig.\ref{fig2}. By computing the gap
equations (\ref{gap1}) and then finding the minimum of the free
energy ${\cal F}$, we find that the superfluid is in homogeneous
state at high temperature and FFLO state at low temperature.
However, the number susceptibility in the region of low temperature
and low number asymmetry is not positive-definite, the FFLO state in
this region is therefore unstable against the number fluctuations,
and the ground state is probably an inhomogeneous mixture of the BCS
superfluid and normal fermion fluid. The shadowed region is the
gapless superfluid with $\delta\mu > \Delta$ where the energy gap to
excite qusi-particles is zero and the system may be sensitive to the
thermal and quantum fluctuations. The phase diagram with both inter-
and intra-species pairings is shown in the right panel of
Fig.\ref{fig2}. Besides the familiar phase with only inter-species
pairing ($\Delta\ne 0, \Delta_a=\Delta_b=0$) and the expected phases
with only intra-species pairing ($\Delta_a, \Delta_b\ne 0, \Delta=0$
and $\Delta_b\ne 0, \Delta=\Delta_a=0$), there appears a new phase
where the two kinds of pairings coexist ($\Delta, \Delta_a,
\Delta_b\ne 0$). In this new phase the FFLO momentum is zero and the
number susceptibility is negative, $\chi < 0$. Therefore, the
homogeneous superfluid in this region is unstable against number
fluctuations, and the ground state is probably a inhomogeneous
mixture of these three superfluid components. In the familiar phase
with only inter-species pairing, there remain a stable FFLO region
and an unstable FFLO triangle where the number susceptibility is
negative and the system may be in the state of phase separation. As
we expected in the introduction, the gapless state appears only in
the inter-species pairing superfluid, and in the region with
intra-species pairing all the fermions are fully gapped.

In summary, we have investigated the phase structure of an
asymmetric two-species fermion superfluid with both inter- and
intra-species pairings. Since the attractive interaction for the
intra-species pairing is relatively weaker, its introduction
changes significantly the conventional phase diagram with only
inter-species pairing at low temperature. For systems with fixed
chemical potentials, the inhomogeneous superfluid with
inter-species pairing at low temperature is replaced by the
homogeneous superfluid with intra-species pairing, while for
systems with fixed species numbers, the two kinds of pairings can
coexist at low temperature and low number asymmetry. In any region
with intra-species pairing, the interesting gapless superfluid is
washed out and all fermion excitations are fully gapped. To
finally determine the exact state of the system in the regions
with negative number susceptibility ($\chi < 0$) in
Fig.\ref{fig2}, further investigation, especially considering
other possible inhomogeneous superfluid phases, is needed.

Since the intra-species pairing is significant at low temperature,
our result including both inter- and intra-species pairings is
expected to influence the characteristics of those low temperature
fermion systems. For instance, it will change the equation of
state of the nuclear superfluid (where neutron-proton and
neutron-neutron, proton-proton pairings play the roles of inter-
and intra-species pairings) or the color superconductor of quark
matter (where up-down and up-up, down-down quark pairings play the
roles of inter- and intra-species pairings in two flavor case) in
compact stars. On the other hand, systems involving two different
alkali atoms can be used to study the result we obtained.

{\bf Acknowledgments:} We thank L.He, H.Hu, M.Huang and M.Jin for
helpful discussions. The work was supported by the grants
NSFC10575058, 10428510 and 10435080.


\begin{thebibliography}{20}
\bibitem{chi} Note that $\chi_{ij}=dn_i/d\mu_j$ is always positive for uniform systems and
$\chi<0$ is a signature of a phase separation.
\end{thebibliography}
\end{document}